\documentclass[aps,prb,twocolumn,showpacs,superscriptaddress,groupedaddress,10pt]{revtex4}
\usepackage{amsmath}
\usepackage{amssymb}
\usepackage{graphicx}
\usepackage{color}
\usepackage{epsfig}
\usepackage{dcolumn}
\usepackage{bm}
\usepackage{bm}
\usepackage{blindtext}
\usepackage{natbib}

\def\be{\begin{equation}} \def\ee{\end{equation}}
\def\bea{\begin{eqnarray}} \def\eea{\end{eqnarray}}

\newcommand\mc            {\mathcal}
\def\eps{\varepsilon}

\begin{document}
\title{Quantum critical dynamics
  for a prototype class of insulating antiferromagnets}
\author{Jianda Wu}
\affiliation{Department of Physics, University of California, San
Diego, California 92093, USA}
\author{Wang Yang}
\affiliation{Department of Physics, University of California, San
Diego, California 92093, USA}
\author{Congjun Wu}
\affiliation{Department of Physics, University of California, San
Diego, California 92093, USA}
\author{Qimiao Si}
\affiliation{Department of Physics and Astronomy and Center for
Quantum Materials, Rice University, Houston, Texas 77005, USA}

\begin{abstract}
Quantum criticality is a fundamental organizing principle for studying
strongly correlated systems.
Nevertheless, understanding quantum critical dynamics at nonzero
temperatures is a major challenge of condensed matter physics due
to the intricate interplay between quantum and thermal fluctuations.
The recent experiments in the quantum spin dimer material TlCuCl$_3$
provide an unprecedented opportunity to test the theories of quantum
criticality.
We investigate the nonzero temperature quantum critical spin dynamics
by employing an effective $O(N)$ field theory.
The on-shell mass and the damping rate of quantum critical
spin excitations as functions of temperature are calculated
based on the renormalized coupling strength,
which are in excellent agreements with experiment observations.
Their $T\ln T$ dependence is predicted to be dominant at very
low temperatures, which is to be tested in future experiments.
Our work provides confidence that quantum criticality as a theoretical
framework, being considered in so many different contexts of condensed
matter physics and beyond, is indeed grounded in materials and
experiments accurately.
It is also expected  to motivate further experimental investigations on
the applicability of the field theory to related quantum critical systems.
\end{abstract}
\maketitle

\section{Introduction}
\begin{figure}[h!]
\begin{center}
\includegraphics[width=0.85\columnwidth]{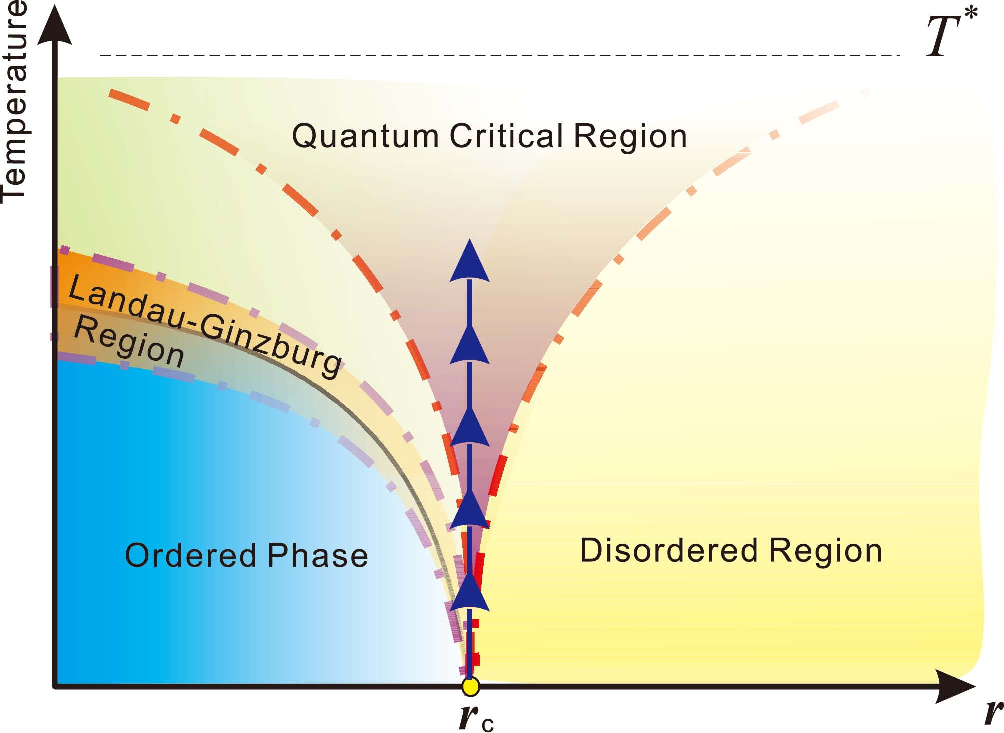}
\end{center}
\caption{{Above figure depicts a generic phase diagram with a second order quantum phase transition, which arises in broad classes of systems \cite{Millis1993,Si_Science2010,Ruegg2014,Schuberth2016,Wu2016b}. The $r$ and $r_c$ denote tuning parameter and quantum critical point respectively. Due to different correlation characters resulting from the competition between thermal and quantum fluctuations, the whole diagram has been {\it{de facto}} divided into different regions as it is manifestly illustrated with different colors. The solid curve is used to denote a second order classical phase transition while the dot-dash curves illustrate crossover regions qualitatively. The broken line on the top is used to denote the cutoff temperature, $T^*$, beyond which quantum criticality is negligible. The dark blue arrow, flowing right upward from the QCP, shows the region we study in the main text.}}
\label{typical}
\end{figure}

Quantum and thermal fluctuations combine to determine the overall
nonzero-temperature quantum dynamics as well as thermodyanmics
of strongly correlated many-body systems.
A quantum critical point 
occurs at zero temperature when matter goes from one quantum
ground state to another upon tuning a non-thermal parameter
\cite{SpecialIssue2010,Sachdev2011a}.
An illustration of a generic phase diagram of quantum phase
transition is presented in Fig. \ref{typical}.
Physical properties around quantum critical points are of extensive
current interest.
For instance, quantum criticality gives rise to unusual spin dynamics
in heavy fermion metals \cite{Si2001,Schroder2000,Schuberth2016}
and one-dimensional quantum magnets \cite{Kinross2014,Wu2014},
as well as the non-Fermi liquid behavior and unconventional
superconductivity in a variety of strongly correlated electron systems
\cite{SpecialIssue2010,Coleman2005,Si_Science2010,Wu2016b}.
The corresponding real-frequency dynamics is in general difficult
to calculate, especially at nonzero temperatures $(T > 0)$.
Indeed, even for quantum systems in one spatial dimension, analytical
understandings of such dynamical properties are still limited
\cite{Wu2014,Deift1994,Sachdev1997,Caux2006,Caux2011}.

Recently, the neutron scattering experiments in the three-dimensional
quantum magnetic dimer compound TlCuCl$_3$ have provided an ideal
testbed of quantum
critical theory at an unprecedented level \cite{Ruegg2004,Ruegg2005,
Ruegg2008,Ruegg2014}.
TlCuCl$_3$ undergoes a continuous quantum phase transition from the
quantum disordered phase to the N$\acute{\rm{e}}$el ordered one
with increasing pressure.
At ambient pressure, the ground state is a dimerized singlet paramagnet
with the gapped low-energy triplon (triplet state) excitations.
Upon increasing pressure, the triplon band bottom is lowered, and,
at a critical pressure, these excitations become gapless leading
to a quantum phase transition into the N\'eel ordered ground state.
Correspondingly, the low-energy phase is capatured by a generic
3 (space) + 1 (time) dimensional relativisitic
O(3) quantum $\phi^4$ theory
\cite{Chakravarty1989,Chubukov1994} [Appendix \ref{app:lagrangian}].
While a great deal of efforts, such as theoretical proposals for detection of the Higgs mode and the ratio of gaps in disorder and order regimes,  have been made towards understanding the
experiments \cite{Kulik2011,Jin2012,Chen2013,Scammell2015,
Witczak-Krempa2015,Meng2015},
the dynamics in the quantum critical regime have not been understood
properly \cite{Scammell2017}.

In this article, we study the critical dynamics of quantum antiferromagetism
at nonzero temperatures by employing a generic 3 (space)+1 (time)
dimensional relativistic $O(N)$-invariant quantum $\phi^4$ theory.
Both the on-shell and off-shell quantum critical dynamics at $T>0$
are calculated in a broad dynamical regime.
Our on-shell results for the mass and damping rate allow not only
a qualitative understanding of the experimental observation but also
a quantitative description of their magnitudes, and the $T\ln T$
dominance is found at very low temperatures.
Moreover, the effective coupling of the field theory appropriate to
systems such as TlCuCl$_3$ is determined, which will be important
for further experimental means to test the applicability of
the field theory to these materials.
Studying the material and the quantum field theory serves as a means
to explore the Higgs physics in a tabletop setting \cite{Higgsontable}.

\section{The Model}

We start with the following $D$-dimensional
($D=d+1$) Euclidean relativistic Lagrangian with $N$-components $O(N)$ real field
$(\phi_1 ,\phi_2 , \cdots ,\phi_N )$ [Appendix \ref{app:lagrangian}],
\be
{\mc{L}}_0  =
\frac{1}{2} {\left( {\partial _\nu  \phi _i
(\tau ,\mathord{\buildrel{\lower3pt\hbox{$\scriptscriptstyle\rightharpoonup$}}
\over x} )} \right)^2 }  + \frac{{g^2 }}{4!}\mu ^{2\eps }
{\phi _i^2 (\tau ,\mathord{\buildrel{\lower3pt\hbox{$\scriptscriptstyle\rightharpoonup$}}
\over x} )\phi _j^2 (\tau ,\mathord{\buildrel{\lower3pt\hbox{$\scriptscriptstyle\rightharpoonup$}} \over x} )},
\label{Lagrangian0}
\ee
where the mass is set as zero at zero temperature corresponding to the
quantum critical point,
 $\partial_\nu= ({\partial _\tau,
\partial_{\mathord{\buildrel{\lower3pt\hbox{$\scriptscriptstyle\rightharpoonup$}}\over x}} } )$,
and $\mu$ is the energy scale parameter.
In addition,
$g$ is the dimensionless coupling constant at the energy scale of $\mu$.
The ultraviolet (UV) divergent terms in the renormalization process
will be absorbed by counter terms order by order in the framework
of dimensional regularization (D-Reg), and $\eps$ is taken as
zero for comparing theoretical results with experimental measurements.
At the quantum critical point, the mass is renormalized to zero by definition.
At nonzero temperatures, the mass only depends on temperature,
therefore, the region we shall investigate is the one illustrated
by the blue arrows in Fig.~\ref{typical}.
The counter terms are not displayed but will be determined
following  the minimal-subtraction scheme \cite{Wit1986}.

We study $\mc{L}_0$ for the dynamics at nonzero temperatures within the
Braaten-Pisarski resummation program, which is a systematic calculation
machinery taking advantages of D-Reg and resummation
\cite{Pisarski1988,Pisarski1989,Braaten1990,Pisarski1991,Andersen2005}.
The relevant diagrams up to 2-loop contributions are shown in
Figs.~S1 (a--h) in Appendix B.
The finite-temperature renormalized theory at $N=1$ was obtained previously
in Refs. \cite{Banerjee1991,Parwani1992,Parwani1995,Wang1996,Laine2015}.
We consider the field theory with a general $N$, and
the case of TlCuCl$_3$ corresponds to $N=3$.
Starting from $\mc{L}_0$, a 1-loop calculation for the self-energy
or, the thermal mass, is based on the diagram of Fig.~S1(a) in Appendix B,
giving rise to
\be
m^2_T=-\Sigma_0(q)=a_N g^2(\mu )T^2
\ee
with $a_N = (N+2)/72$.
By using the thermal mass, an effective field theory is constructed
\be
\mc{L}_2  = \frac{1}{2}[(\partial _\mu  \phi _i )^2  + m_T^2 \phi _i^2 ]
+ \frac{1}{4!}g^2\mu ^{2\varepsilon } \phi _i^2 \phi _j^2  - m_T^2 \phi _i^2/2
\ee
with a new vertex $-\frac{1}{2}m_T^2 \phi _i^2$ illustrated by Fig.~S1(b)(Appendix B).
$\mc{L}_2$ naturally cures the infrared (IR) divergence, equivalent
to resumming the daisy diagrams (Fig.~S2 in Appendix B
based on $\mc{L}_0$ \cite{Laine2015}.
This procedure leads to a non-analytic (in $g^2$) 1-loop renormalized mass,
\be
{m'}_T^2  = m{_T}^2  - 3m_T^3 /(\pi T) + O(g^4 \ln g),
\ee
showing the non-perturbative nature of the quantum $\phi^4$
theory \cite{Parwani1992,Laine2015,Schwartz2013}.
In order to determine the quantum dynamics, we proceed to the 2-loop level.
Recognizing the 1-loop correction for the interaction (Fig.~S1(c)
Appendix B),
a 1-loop zero-temperature interaction counter term
(Fig.~S1(d) in Appendix B)
is incorporated to obtain the full 1-loop renormalized effective
field theory $\mc{L}_3$,
\bea
\mc{L}_3  &=& \frac{1}{2}\left[ {\left( {\partial _\mu  \phi _i }
\right)^2  + {m'}_T^2 \phi _i^2 } \right]
+ \frac{{g^2 }}{4}\mu ^{2\varepsilon } \phi _i^2 \phi _j^2 \nonumber \\
&& - \frac{1}{2} {m'}_T^2 \phi _i^2 + \frac{{g^2 \mu ^{2\varepsilon } }}{{4!}}
\left[ {\frac{{N + 8}}{6}\frac{{g^2 }}{{(4\pi )^2 }}
\frac{1}{\varepsilon }\phi _i^2 \phi _j^2 } \right]
\eea
which serves as the starting point for further calculations
at the two-loop level.

From $\mc{L}_3$, via Figs.~S1(a,b,e,f,g,h) (Appendix B), we obtain the
renormalized self-energy to the order of $g^4$,
$\Sigma_2=\Sigma '_2  + i\Sigma ''_2$.
The detailed expressions of $\Sigma_2^\prime$ and $ \Sigma''_2$
are presented in Appendix C.
For the long-wavelength physics, we fix the momenta for the external legs to
be zero in the sunset diagram (Fig.S1(h) in Appendix B).
The divergent terms associated with $\varepsilon \to 0$ are absorbed by
counter terms for the wavefunction and momentum renormalizations,
respectively.
The detailed renormalzation calculation up to the two-loop level is given
in Appendix C.
The dynamical structure factor (DSF) is related to the
self-energy through,
\bea
S(\omega ,\vec p = 0)
=\frac{2 {\rm{Im}}[\chi (\omega , {\vec p} = 0)]}{1 - e^{ - \beta \omega } },
\eea
where $\chi^{-1} (\omega ,\vec p = 0) = - \omega ^2  + m_T^{'2}
- \Sigma _2 (\omega )$, then the damping rate follows
$\gamma = \Sigma''_2/(2\omega)$.
In the following we will present the on-shell mass, and damping rates in
three different frequency regimes.
The comparison with the experiments on TlCuCl$_3$
will be discussed.


\section{On-shell dynamics and the damping rate}
A complex pole $\omega = M - i \gamma$ can be obtained from the zeros of
$\chi^{-1}(\omega ,\vec p = 0)$.
Up to the two-loop level, the on-shell mass $M$ is determined as
\bea
M^2(T)  &=& a_N T^2 g^2(\mu) \left\{ { {1 + b_N g^2 (\mu )\ln (T/\mu )}  + } \right.
 \nonumber \\
&& \left. { + c_1 g(\mu ) + c_2 g^2 (\mu )\ln g^2 (\mu ) + c_3 g^2 (\mu )} \right\},
 \label{twoloopmass}
\eea
where $b_N  = (N + 8)/(48\pi ^2 )$, $c_1  =  - 3a_N^{\frac{1}{2}} /\pi ,\;c_2  =
1 /(8\pi ^2 ),\;c_3  = \{  - (b_N/2 ) \ln (4\pi ) + \ln a_N /(8\pi ^2 )
+ [\frac{3}{2}a_N \delta _0  + \delta _1 /(16\pi ^2 )]\}$
with $\;\delta _0  = 5.242$, and $\delta _1  = 3.644$. In addition the $\ln T$ dependence in the big brackets arises from
the sunset diagram (Eq.~(C3) in Appendix C),
signaling physics beyond scaling ansatz as will be discussed later.
Around the complex pole, the renormalized propagator is well
approximated by
\bea
\chi^{-1} (\omega  ,\vec p = 0)
= - \omega ^2  + M^2  - i \Sigma '' _2 (\omega ) .
\eea
Substituting the expression of $\Sigma''_2$ in Appendix C,
the on-shell damping rate reads
\be
\gamma (\omega^2=M^2,\vec{p} = 0)
= g^2 m_T/(64\pi). \label{daming}
\ee

In fact, the renormalized mass $M^2(T)$ in Eq.~(\ref{twoloopmass})
is independent on the choice of the energy scale $\mu$.
To verify this at the two-loop level, we employ the $\mu$-dependence of
coupling strength $g(\mu)$ by solving the one-loop renormalization
group equation \cite{Frenkel1992,Arnold1994,Schwartz2013},
\bea
g^2(\mu )  = \frac{g^2(\mu_0)} {1 + b_N g^2(\mu_0) \ln (\mu_0/\mu )} .
\label{running_coupling}
\eea
$\mu_0$ is a reference energy scale, which for convenience is set at
$\mu_0=T$ throughout the rest part of the paper \cite{FootNote}.
From Eq.~(\ref{running_coupling}), to the order of $g^4$,
we arrive at
\be
g^2 (\mu_0=T) = g^2 (\mu )[1 + b_N g^2 (\mu )\ln (T/\mu )]
+ O(g^6(\mu)),
\ee
which apparently pushes the $\mu$-dependence of
the two-loop mass Eq.~(\ref{twoloopmass}) to the next order $g^5\ln g$.

While our theoretical renormalization procedure is general, the strategy
for comparison with the experimental data is as follows.
We start with a particular temperature $T=T_1$, set as $\mu_0$.
The value of $g(\mu_0=T_1)$ is fitted by comparing the
thermal mass Eq.~(\ref{twoloopmass}) in which $\mu$ is set
as $\mu_0$ with the experimental data at $T=T_1$.
For data measured at a different temperature, say, $T_2$,
Eq.(\ref{running_coupling}) is used to determine $g(\mu=T_2)$,
based on which, the mass and damping rates at $T_2$ are
calculated and compared with the experimental data.
As will be clear later, excellent agreements with experiemental
results are achieved for independent data points of
masses and damping rates.


\begin{figure}[t!]
\begin{center}
\includegraphics[width=7cm]{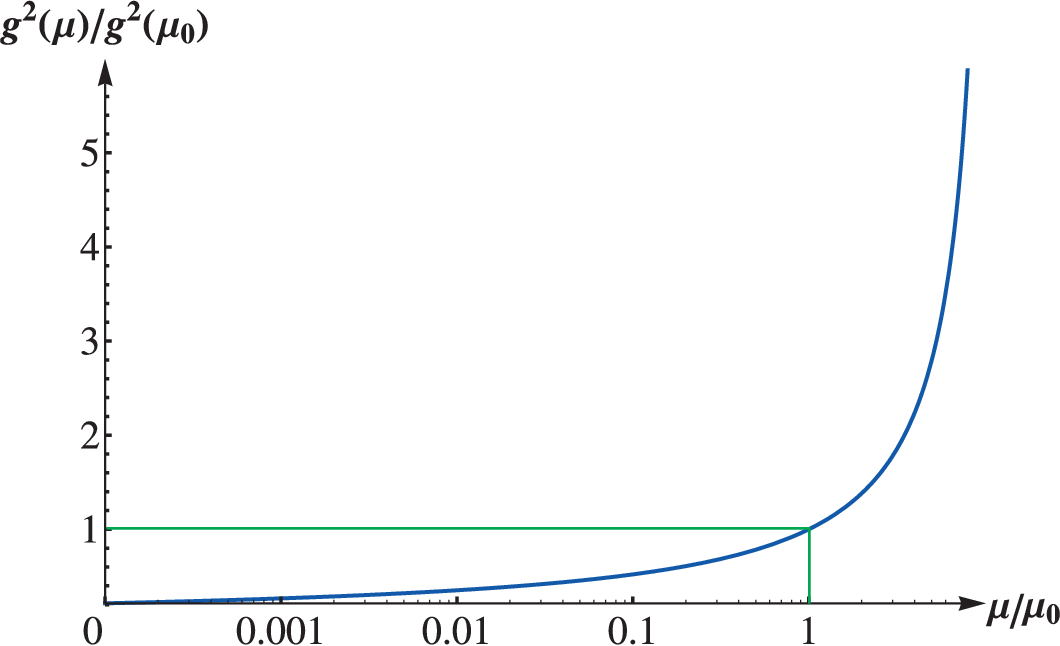}
\end{center}
\caption{Coupling strength at different physical energy scale, $\mu$,
following Eq.~(\ref{running_coupling}).
Here, the reference coupling $g(T) = 4.146$, and the Landau
pole occurs at $\mu_L = 12.253 T$.}
\label{flow}
\end{figure}

\begin{figure}[t!]
\begin{center}
\includegraphics[width=6.8cm]{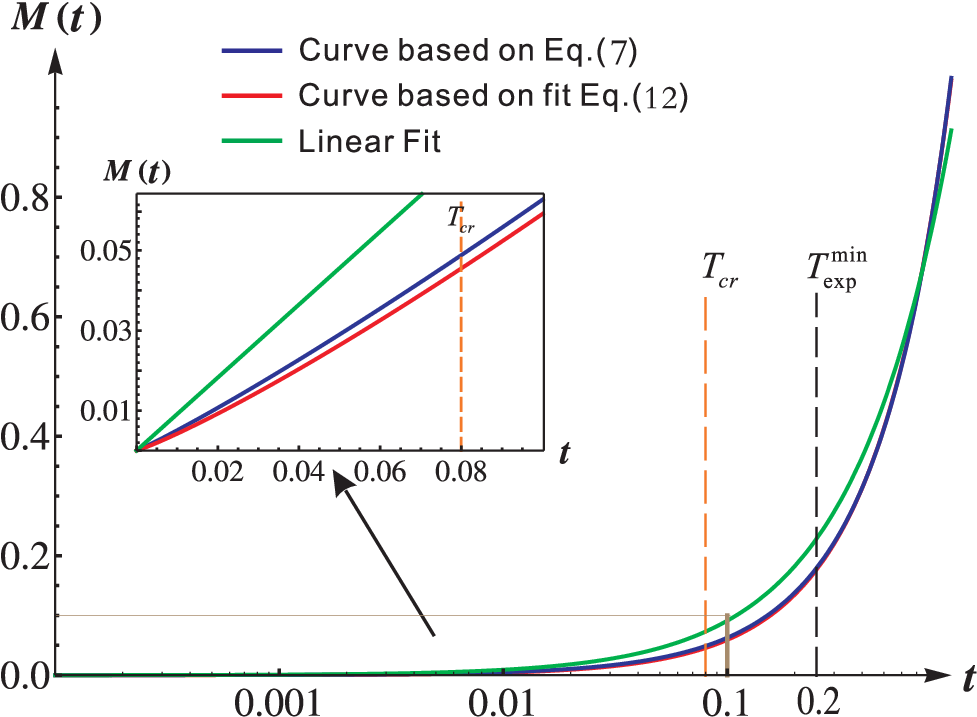}
\end{center}
\caption{Two-loop renormalized mass
{\it{v.s.}} reduced temperature $t$. The blue curve is based on Eq.~(\ref{twoloopmass}). It is approximated by Eq.~(\ref{approxM}) and $M(T) = 0.91 t$ illustrated by red and green curves, respectively. The black and orange broken lines respectively mark the minimal experimental temperature, and the crossover temperature when the $T \ln T$ behavior starts to become dominant. The inset figure zooms in the low temperature region ($t < 0.1$), which clearly shows that the mass behavior significantly deviates from the linear-$T$ behavior.}
\label{masslog}
\end{figure}

In principle, the temperature for any data point can be served as
a renormalization reference point.
For later convenience, we choose maximal temperature,
$T=T_{\rm{exp}}^{\rm{max}}\approx 12K$, at which the data was measured.
Correspondingly, based on Eq.~(\ref{twoloopmass}), the initiative
$g(\mu_0=T_{\rm{exp}}^{\rm{max}})\approx 4.15$ by substituting $N=3$, and
Fig.~\ref{flow} illustrates the flow of $g(\mu)$ at different energy scales.
The advantage of this choice is that the coupling constant $g(\mu)$
determined at lower energy scales becomes smaller, which improves the
accuracy of the perturbative calculation.
In particular, the validity of the perturbation theory can only
be justified at $\mu$ far less than the scale of Landau pole.
Our calculation shows that the Landau pole based on
Eq.~(\ref{running_coupling}) is located at the energy scale about
$\mu_L=12.25  T_{\rm{exp}}^{\rm{max}}$, which is well beyond the
temperature scope measured.

Following this procedure, we plug the obtained coupling constant at
each temperature into Eq.~(\ref{twoloopmass}) to obtain the two-loop
renormalized mass as plotted in Fig. \ref{masslog}.
For later convenience, the calculated masses can be approximated by
(the mass here is reduced by $T_{\rm{exp}}^{\rm{max}}$)
\be
M(T)= 1.00 t/[1.00 + 0.30 \ln (1/t)] \label{approxM}
\ee
with $t = T/T_{\rm{exp}}^{\rm{max}}$.
The linear-$T$ behavior dominates at relatively high temperature
in which the experiment measurements were performed.
As lowering the temperature, the $T\ln T$ correction becomes important
and dominates in the very low temperature region.
This expression can be understood as follows: By setting $\mu=T$ in
Eq.~(\ref{twoloopmass}), the leading contribution to $M(T)\sim Tg(\mu=T)$.
The deviation from the linear-$T$ behavior of $M(T)$ is due to the
weakening of $g(\mu=T)$ as lowering $T$ according to the running
coupling constant expression Eq.~(\ref{running_coupling}).
The corresponding crossover temperature scale can be determined as
\bea
T_{cr}\sim \mu_0 e^{-1/[b_N g^2(\mu_0)]}
\eea
with $\mu_0=T^{\max}_{\exp}$.
In our case, $T_{cr}\approx 1$K corresponding to $t\approx 0.08$,
which is about half-order smaller than the lowest temperature in experiments.
Therefore in the range of experiment measurement temperatures Eq.~(\ref{approxM})
gives a theoretical prediction of mass $M_{\rm{th}} \approx 1.00 T$ which excellently agrees
with the experimental result $M_{\exp} = T$ \onlinecite{Ruegg2014}.
However, for the material of TlCuCl$_3$, there exists a small gap
induced by anisotropy $\Delta \approx 0.38$meV \cite{Ruegg2008},
which spoils the $O(3)$ invariance.
Since this temperature scale ($\Delta$) is much larger than $T_{cr}$,
we are not optimistic at the experimental observation of the
$T\ln T$ behavior in the material of TlCuCl$_3$. However, we
expect that the $T \ln T$ behavior should be observed for the material
which can be effectively described by the $\mc{L}_0$ [Eq.~(\ref{Lagrangian0})]
at low enough temperatures.

\begin{figure}[t!]
\begin{center}
\includegraphics[width=6.5cm]{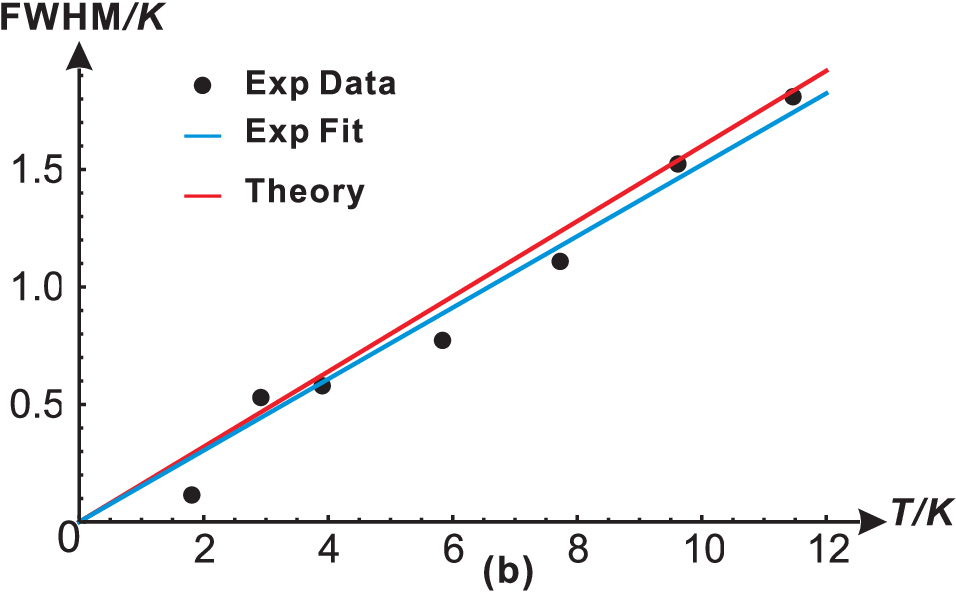}
\end{center}
\caption{Temperature dependence of the damping rate.
Here, our theoretical result ${\rm{FWHM}} = 0.16 T$ (red line)
is shown along with the experimental result of Ref.~\cite{Ruegg2014},
which includes the values (without error bars) at different temperatures (black dots)
and a fit of these values in terms of ${\rm{FWHM}} = 0.15 T$ (blue line).}
\label{dampingevolution}
\end{figure}

We proceed to analyze the damping rates in the experimental relevant
region,  namely, the linear-$T$-dominant region in Fig.~\ref{masslog}.
After some calculations, we arrive at $\gamma_{\rm{th}}
(\mu =T) \approx 0.08 T$ for $N=3$.
Correspondingly, the full width at half maximum (FWHM) scales as
FWHM$_{\rm{th}}\approx 0.16T$
which excellently agrees with the experimental observation of
FWHM$_{\rm{exp}}\approx 0.15T$ \cite{Ruegg2014} (Fig.~\ref{dampingevolution}).
Our calculation is exact at two-loop, $g^4$. However, from Eq.~(\ref{daming}),
it is observed that the mass at the one-loop level ($g^2$) has already resulted
in the damping rate at the
order of $g^4$ (two loops). Nevertheless, the mass will exhibit the logarithmic
dependence at low temperatures, therefore,
we still expect the linear-$T$ behavior for the on-shell damping rate will
also be spoiled when temperature is decreased further to the $T \ln T$
dominant region, which should be verified when the field-theory approach
is pushed to the order of $g^6$ (three loops).

The striking agreement between our theoretical results and the experiment
indicates that the $3+1$ dimensional $O(3)$ relativistic quantum field
theory takes an excellent account of the underlying physics near the
pressure-induced quantum critical point of TlCuCl$_3$.
Furthermore, for the initiative $g$, we have $g^2/4! \approx 0.7$.
This value implies that the interaction among the critical modes is
significant but still moderate, allowing for a loop expansion to
extract semi-quantitative results for the pertinent physical properties.
Furthermore, one must proceed to at least two loop ($g^4$) to
obtain a physical prediction of the $T \ln T$ dominance. Following our two-loop renormalization analysis we
expect a clear $T \ln T$-dominant behavior for the mass at low
temperatures, which goes well beyond conventional scaling ansatz.



\section{Off-shell dynamics at frequencies far away from $M$}
We first determine the damping rate in the zero frequency limit,
$\omega \ll M$.
The imaginary part of the self-energy $\Sigma^{''}_2(q=0,\omega)$
can be organized into the imaginary part of $G_0$, $G_1$ and $G_2$ as
shown in Eqs.~(S6-S9) in Appendix C, which arise from the contribution of
of the sunset diagram (Fig.~S1(h) in Appendix B).
In the low frequency limit, due to the on-shell
energy-momentum conservations, there is no contribution from $G_0$
because of a three-particle threshold.
The phase space that satisfies the on-shell constraints lies in
the large momentum regime.
Because of the Bose-distribution factor, the contributions from $G_1$
and $G_2$ are exponentially suppressed.
Our calculation shows that
\bea
\gamma &\sim& T^2 \exp[-(m_T/T)(m_T/\omega)]/(2\omega) \nonumber \\
&\sim& T
\exp[-g^2 T/\omega)]/(2(\omega/T)) \to 0 \;\; (\omega \to 0).
\eea
This exponentially small result is subleading compared to the
perturbative renormalized theory we carried out at the order of $g^4$.
As a consequence of the suppressed spectral weight in this regime,
the analysis is beyond the scope of the procedure outlined here.

\begin{figure}[t!]
\begin{center}
\includegraphics[width=7cm]{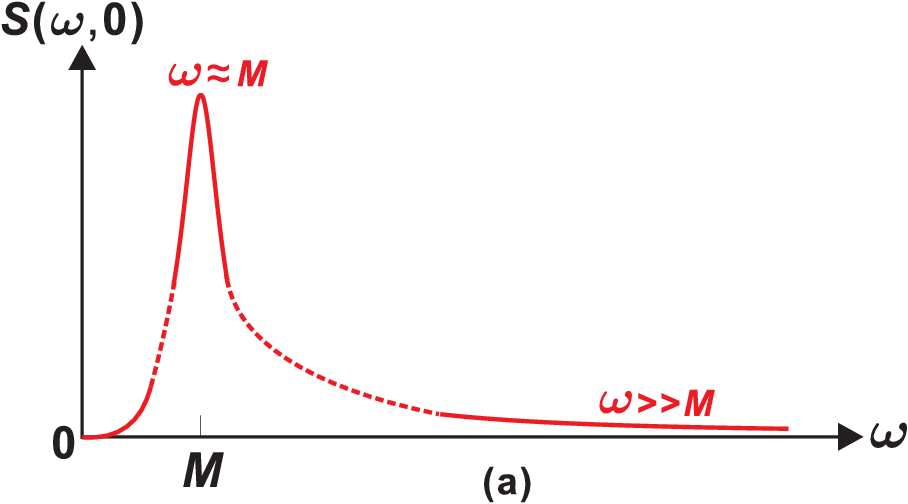}
\end{center}
\caption{A schematic plot of the dynamical structure factor at zero
momentum, describing the behavior in different asymptotic regimes:
on-shell ($\omega\sim M$) and off-shell at $\omega \ll M$ and $\omega \gg M$.
The dash lines interpolate among these regimes.}
\label{dynamics}
\end{figure}

We next consider the damping rate in the large frequency limit, $\omega \gg M$,
for which the physical energy scale should be set by $\mu = \omega$.
On the other hand, the perturbative framework stops working at the
energy scale of the Landau pole
$\mu_L=\exp \left\{ {1 /[b_N g(T)^2 ]} \right\}$ determined
by Eq.~(\ref{running_coupling}).
Thus, we work in the range $M(\mu = \omega) \ll \omega \ll \mu_L$.
In the large frequency limit, the dominant contribution to $\Sigma''_2$
comes from ${\rm{Im}} G_0$. Evaluating the integration for ${\rm{Im}} G_0$ exactly yields ${\mathop{\rm Im}\nolimits} G_0 (\omega ^2 ,\vec p = 0)
\sim g^4(\omega) \omega^2$ (Appendix C), correspondingly, $\gamma \sim g^4(\omega) \omega$ and $S(\omega ,\vec p = 0) \sim \omega^{-2}$.
In this regime, the system is over-damped.
A logarithmic correction naturally arises from $g^2(\omega)$
determined from Eq.~(\ref{running_coupling}), but is of the order $g^6$.
The overall behavior of the DSF is shown in Fig.~\ref{dynamics},
which we expect to be experimentally observed in the near
future in proper strongly correlated systems.

\section{Discussions}
We note that any physical observable should be cut-off independent.
Our results satisfy this requirement at the two-loop level, while, in
contrast, previous results along these lines are cut-off independent
only at the order of one loop \cite{Scammell2017}.
Improvement is significant and
fundamental rather than technical, since it brings qualitatively new
behavior of the mass as well as the dynamic information: The $T \ln T$
behavior beyond the scaling ansatz emerges only at the two-loop level;
it cannot be accessed by the method of [\onlinecite{Scammell2017}],
being correct at one loop
only and accurate up to the order of $g^3$ (or $\alpha^{3/2}$ in the
notations of that study); accordingly, the two-loop RG invariant
mass as presented in Eq.~(\ref{twoloopmass}) here does not obtain by the method of
[\onlinecite{Scammell2017}]. There are also differences between our study and
that of [\onlinecite{Scammell2017}] so
far as the potential for guiding experiments is concerned, and we
believe that, also here, there are considerable advantages in using
our approach.

We now remark on a few points.
First,the two-loop calculation
has incorporated all the pertinent terms at the order of $g^4$.
There are an infinite number of other diagrams which can contribute
to the order of $g^4$; for example, if adding one more bubble in
Figs.~S1(e,f,h) (Appendix B) or one more blob (two-point interaction vertex)
in the Fig.~S1(h), their contribution is at the order of $g^4$.
However, when summing over all of these kinds of diagrams,
their contributions at the order of $g^4$ exactly cancel with
each other \cite{Parwani1992}, leaving a final contribution at
the three-loop level.
The procedure can in principal be carried out order by order, leading
to any desired accuracy for the (on-shell) quantum critical dynamics
at nonzero temperatures of the relativistic 3+1 dimensional
$O(N)$ quantum $\phi^4$ theory \cite{Andersen2005}.
In our study, the result at the two-loop level is already in an
excellent agreement with the experiments.
We have also demonstrated that the effective coupling constant
provides a justification for the two-loop calculation.

Second, our analysis is asymptotically exact for $D=3+1$ dimensions,
where all the UV divergences are systematically absorbed by counter
terms controlled by the small quantity $\eps$.
These UV divergences associated with $\eps$ yield proper wave-function
and momentum renormalizations order by order in the form of counter
terms \cite{Parwani1992}, which in turn modify the next-leading-order behavior.
In other words,  $\eps$ can only be taken to zero when, at each order,
the contributions from the previous-order counter-terms have been incorporated.
For lower-dimensional systems, say, in two spatial dimensions,
the corresponding problem can be analyzed by setting $\eps = 1/2$.
We want to emphasize the difficulty in removing the IR singularities.
Not in every field theory can one remove its IR singularities
\cite{Laine2015}.
It is fortunate that there is a systematic way to regularize the $3+1$
dimensional quantum $\phi^4$-theory
However, the resummation in the calculation paradigm is essentially
non-perturbative, which gives rise to non-analytical corrections
beyond the perturbation series, therefore, the procedure can become
very involved when one tried to push to higher orders.

Third, the leading behavior we obtained for TlCuCl$_3$ is expected,
from the perturbation view point, to hold when $|m^2(T = 0)|\ll T$
(Fig.~\ref{typical}, also called as high-$T$ regime
\cite{Sachdev2011a}).
However, when one moves to the regime $|m^2(T=0)|\gg T$ (also called as
low-$T$ regime \cite{Sachdev2011a}) the non-critical background does
become important.
For obtaining a correct understanding one not only needs to introduce new
counter terms associated with the open gap to cure newly appeared UV
divergences, but also require new techniques beyond the current approach
to access the physics in the ordered regime especially near the
transition line \cite{Pietroni1998}.
These new must-introduced techniques and the expected new physics
compose a new challenging goal worthy of further exploration.

\section{Conclusions}
We have studied the quantum critical dynamics at nonzero temperatures
based on a generic 3+1 dimensional relativistic quantum $\phi^4$ theory
and compared the results with the spin dynamics experiments on
the insulating antiferromagnet TlCuCl$_3$.
The theoretical coupling constant is determined from the temperature
dependence of the thermal mass corresponding to the gap value.
The resulting mass and damping rate are consistent with the experimental data.
The logarithmic behavior is predicted to be dominant at extremely
low temperature which should be observable in a
proper experimental setup.
At last, the damping behavior over a broad dynamical regime is also
determined for comparison with future experiments.
In turn, our results provide concrete evidence for the applicability
of the underlying quantum field theory to the description of the
observed quantum critical point.
This, together with the quantitative determination of the effective
coupling constant, will allow for the calculation of additional
experimental observable such as the nuclear magnetic resonance (NMR)
relaxation rate.
From a theoretical perspective, knowledge on the dynamical properties
of generic quantum field theory deepens our understandings on
not only condensed matter systems but also the dynamical processes
in high-energy and cosmology physics
\cite{Bednyakov2015}.
Our calculation represents progress
in a theory that is grounded in a concrete and prototype condensed
matter system.


\section{Acknowledgements}
We thank R. R. Parwani for useful discussions, J. H. Pixley
for his contributions during an early stage of the work,
and C. R\"uegg for a motivational discussion.
This work has in part been supported by the
NSF Grant No. DMR-1410375 and AFOSR Grant No. FA9550-14-1-0168,
President's Research Catalyst Award No. CA-15-327861 from the
University of California
Office of the President (J. W.\, W. Y.\, and C. W.),
the NSF Grant No. DMR-1611392 and the Robert A. Welch Foundation
Grant No. C-1411 (J.W.\ and Q.S.).
J. W. acknowledges the hospitality of
Rice Center for Quantum Materials.



\bibliography{higgs}
\bibliographystyle{apsrev-nourl}

\appendix
\setcounter{figure}{0}
\makeatletter
\renewcommand{\thefigure}{S\@arabic\c@figure}


\section{A B\MakeLowercase{rief} D\MakeLowercase{erivation} \MakeLowercase{for} E\MakeLowercase{q.~(1)} \MakeLowercase{in} M\MakeLowercase{ain} T\MakeLowercase{ext}} \label{app:lagrangian}

Each unit cell of TlCuCl$_3$ consists of a dimer formed by two magnetic
spin-$\frac{1}{2}$ Cu$^{2+}$ ions, and the centers of the dimers form
a three-dimensional cubic lattice.
The intra- and inter-dimer antiferromagnetic exchange interactions
are modeled as
\bea
H = J\sum\limits_\textbf{r} {\textbf{S}_{\textbf{r},1}
\cdot \textbf{S}_{\textbf{r},2} }
+ \lambda J\sum\limits_{\left\langle {\textbf{r},{\textbf{r}'}}
\right\rangle;\;l = 1,2} {\textbf{S}_{\textbf{r},l}
\cdot \textbf{S}_{{\textbf{r}'},l} }\;, \label{eq:J1J2}
\eea
where $\textbf{r}$ represents the central position of the dimer,
and $l$ is the site index inside the dimer;
$\textbf{S}_{\textbf{r},1}$ are spin-$\frac{1}{2}$ operators;
$\left\langle \cdots \right\rangle$ denotes summation over the
nearest neighbours.
In addition, $J$ ($>0$) and $\lambda J$ ($0 \leq \lambda \leq 1$) denote
the intra- and inter- dimer coupling strengths respectively.
The next-next-nearest neighbour interactions \cite{Cavadini1999,
Cavadini2000,Cavadini2001,Oosawa2002} are sub-leading and neglected.
When $\lambda = 0$ the ground state is a paramagnetic state with separable
singlet states; when $\lambda = 1$, $H$ describes the cubic lattice
antiferromagnet with a long range N\'{e}el order.
Therefore, a quantum phase transition arises when $\lambda$ is tuned
to a critical value $\lambda_c$.
The observed strong intra-dimer interaction \cite{Cavadini1999,Cavadini2000,
Cavadini2001,Oosawa2002} indicates the dominant low-lying excitations
are spin triplets, therefore, it is proper to introduce bond-operators
$s^\dagger(\mathbf{r})$ and $t_\alpha^\dagger(\mathbf{r}) (\alpha = x,y,z)$
for the dimer spin singlet and triplet states.
They satisfy
$s^\dagger(\mathbf{r}) \left| 0 \right\rangle  =\frac{1}{\sqrt 2}
\left[\left| { \uparrow  \downarrow } \right\rangle_\mathbf{r} -
\left| { \downarrow  \uparrow } \right\rangle_\mathbf{r}\right], \;
t_+^\dag (\mathbf{r})  \left| 0 \right\rangle  =
- \left| { \uparrow  \uparrow }\right\rangle_\mathbf{r},\;
t_z^\dag (\mathbf{r}) \left| 0 \right\rangle
= \frac{1}{\sqrt 2}
\left[ \left| { \uparrow \downarrow }\right\rangle_\mathbf{r}
+ \left| { \downarrow  \uparrow } \right\rangle_\mathbf{r} \right ]$,
and
$t_-^\dag (\mathbf{r})\left| 0 \right\rangle  =
\left| { \downarrow  \downarrow } \right\rangle_\mathbf{r}$
with $ \left| 0 \right\rangle$ being a reference vacuum state
with $t_{\pm}^\dagger = t_x \pm i t_y$ \cite{Sachdev1990}.
In the continuum limit we express the $t_\alpha$-field with
$t_\alpha(\mathbf{r})\sim a^{d/2} \left[\phi_\alpha (\mathbf{r})+
  i \pi_\alpha (\mathbf{r})\right ]$ ($\alpha= x, y, z$)
where $d$ is the spatial dimension and $a$ is the lattice constant.
$\pi_\alpha$ are conjugate momentum to  $\phi_\alpha$ satisfying
$[ \phi_\alpha (\mathbf{r}),  \pi_\beta (\mathbf{r^\prime})]=
i\delta(\mathbf{r}-\mathbf{r^\prime})\delta_{\alpha\beta}$.
Taking advantage of the condensation of the $s$-field, we integrate
over the $\pi$-field, leaving a Lagrangian density in the imaginary-time
formalism as a functional of the $\phi$-field up to quartic
order \cite{Cabra2012},
\be
{\mc{L}}  = \chi (\partial _\tau  \phi _\alpha  )^2
+ \rho _s (\nabla \phi _\alpha  )^2  + m^2 \phi _\alpha ^2
+ u \phi _\alpha ^2 \phi _\beta ^2,
\ee
where the parameters are estimated as
 $\chi \approx 1/J,\; \rho_s \approx 2z a^2 \lambda J, m^2
\approx J(1- 4 z \lambda)$ and $u \approx 2 z a^d  \lambda J$ with
$z$ being the coordinate number,
then the velocity of the field $c_0^2 = \rho_s/\chi \approx
2 z \lambda a^2 J^2$.
After normalizing ${\mc{L}}$ by $\rho_s$, setting $c_0^2 =1,\; m^2 = 0$
and generalizing the field to $N$-components $O(N)$ field
$(\phi_1 ,\phi_2 , \cdots ,\phi_N )$, we recover the Lagrangian $\mc{L}_0$ Eq.~(\ref{Lagrangian0})
presented in the main text.

\section{R\MakeLowercase{elevant} F\MakeLowercase{eynman} D\MakeLowercase{iagrams} \MakeLowercase{up to} T\MakeLowercase{wo} L\MakeLowercase{oops}
\MakeLowercase{and} A T\MakeLowercase{ypical} D\MakeLowercase{aisy} D\MakeLowercase{iagram} \MakeLowercase{for} $\mc{L}_0$ \MakeLowercase{and} R\MakeLowercase{esummation}}

\begin{figure}{t}
\begin{center}
\includegraphics[width=7.5cm]{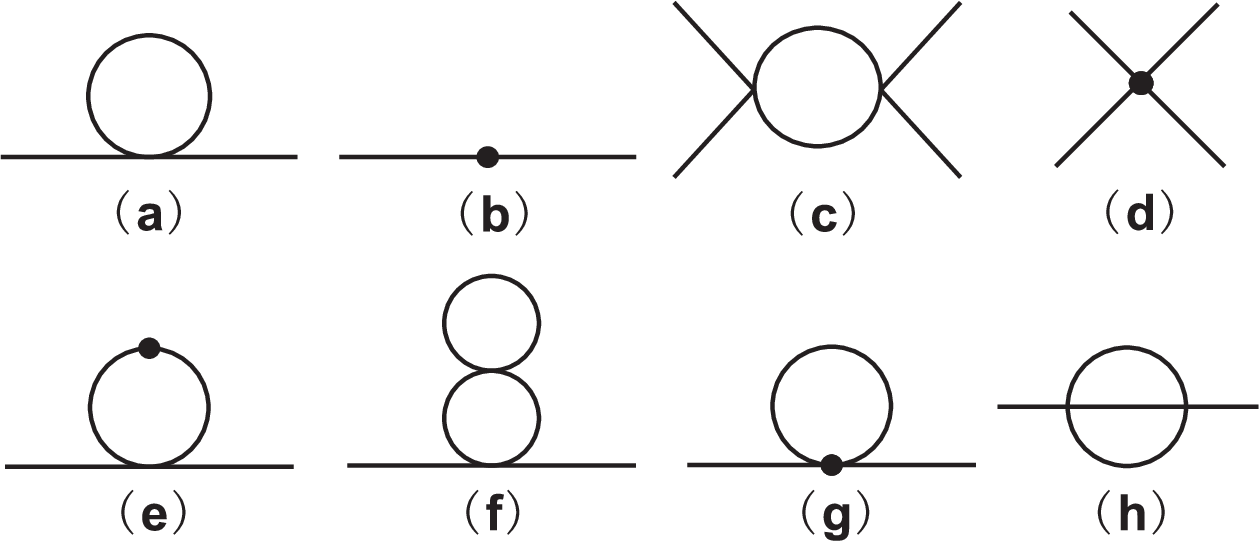}
\end{center}
\caption{(a) One-loop self energy diagram. (b) Two point effective interaction in $\mc{L}_2$.
(c) One-loop correction to the four-point vertex. There are two additional crossed channels, which are not shown.
(d) UV vertex counter term for the interaction.
(e) Insertion of the finite two-point interaction in $\mc{L}_2$ into the one-loop self-energy (bubble) diagram.
(f) Two-loop bubble diagram. (g) Self energy contribution from the UV vertex counter term. (h) Sunset diagram.}
\label{diagrams}
\end{figure}
Figs.~{\ref{diagrams}}(a--h) list relevant Feynman diagrams up to two loops.



\begin{figure}[t]
\begin{center}
\includegraphics[width=6cm]{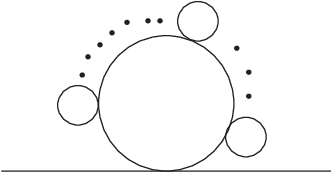}
\end{center}
\caption{A typical daisy diagram with $m$ dressing bubbles for the $\mc{L}_0$ (Eq.~(2) in main text).
}
\label{daisy}
\end{figure}
A typical (one-loop) daisy diagram with $m$ dressing bubbles for the $\mc{L}_0$ (Eq.~(2) in the main text) is illustrated in Fig.~(\ref{daisy}). The ``resummation" of a series of this kind of diagrams helps remove the IR divergence for the $\mc{L}_0$.
Higher order resummation diagrams can be similarly generated when we proceed to multi-loop situations. After adding proper UV counter terms and resummation of all daisy diagrams at each order the Lagrangian can then be updated to the next leading order. In this way it provides a systematic way to renormalize the $3+1$ dimensional quantum $\phi^4$ theory.

\section{R\MakeLowercase{eal} \MakeLowercase{and} I\MakeLowercase{maginary} P\MakeLowercase{arts} \MakeLowercase{of} S\MakeLowercase{elf} E\MakeLowercase{nergy} \MakeLowercase{to} T\MakeLowercase{wo} L\MakeLowercase{oops} }

Up to two loops, the renormalized real part of the self-energy gives,
\be
\Sigma'_2(\omega ^2 ) = \Sigma_{ren}  + F_1  + F_2  + H  \label{self_energy_real}
\ee
with
\begin{widetext}
\bea
\Sigma _{ren}  &=& \frac{{3m_T^3 }}{{\pi T}} + \frac{{3m_T^2 }}{{4\pi ^2 }}
\left( {\frac{{m_T }}{T}} \right)^2 \ln \frac{{4\pi \mu ^2 }}{{T^2 }} + \frac{{3m_T^2 }}{{2\pi ^2 }}\left( {\frac{{m_T }}{T}} \right)^2 \left( {2c_1  - \frac{5}{2} - \frac{{\gamma _E }}{2}} \right) \label{real0} \\
 F_1 &=& \frac{{N + 2}}{3}\frac{1}{{(4\pi )^2 }}\frac{{g^4 T^2 }}{{24}} \left( {\ln \frac{{4\pi \mu ^2 }}{{{m'}_T^2 }} + 2 - \gamma _E } \right) \label{realf1} \\
 F_2  &=& \frac{{N + 2}}{3}\frac{{g^4 }}{{8(2\pi )^4 }}\int_0^\infty  {\frac{{kn_k dk}}{{E_k }}\int_0^\infty  {\frac{{dq}}{{E_q }}
 \left( {q\ln \left| {\frac{{X_ +  }}{{X_ -  }}} \right| - 4k} \right)} } \label{realf2} \\
H &=& \frac{{N + 2}}{3}\frac{{g^4 }}{{8(2\pi )^4 }}\int_0^\infty  {\frac{{kn_k dk}}
{{E_k }}\int_0^\infty  {\frac{{qn_q dq}}{{E_q }}\ln \left| {\frac{{Y_ +  }}{{Y_ -  }}} \right|} } \label{realh}
\eea
\end{widetext}
where ${m'}_T^2  = m_T^2  - 3m_T^3 /(\pi T) + O(g^4 \ln g)$ and $\Sigma_{ren}$ comes from the renormalized contribution of Figs.~{\ref{diagrams}}(a,b,e,f,g). In addition $F_{1}, F_2 (\omega ^2 )$
and $H(\omega ^2 )$ denote those from the real part of the sunset diagram Fig.~{\ref{diagrams}}(h),
in which $c_1  = (2\gamma _E  - 2\ln (4\pi ) - 1)/4$, $X_ \pm   = [\omega ^2  - (E_k  + E_q  + E_{k \pm q} )^2 ]
[\omega ^2  - (E_q  - E_k  + E_{k \pm q} )^2 ]$, and $Y_ \pm   = X_ \pm  [\omega ^2  - (E_k  - E_q  + E_{k \pm q} )^2 ]
[\omega ^2  - (E_k  + E_q  - E_{k \pm q} )^2 ]$ with $E_l^2  = l^2  + m_T^2$, $n_l  = 1/(e^{\beta E_l }  - 1)$, $l = k,q$.
(Here, $\gamma_E$ is the Euler constant.)


The imaginary part of the self energy gives,
\be
\Sigma''_2(\omega^2) = {\rm{Im}} ({G_0}) + {\rm{Im} }({G_1}) + {\rm{Im}}( {G_2}), \label{self_energy_imaginary}
\ee
where
\begin{widetext}
\bea
 G_0  &=& g^4\int {d[k,q]S(E_k ,E_q ,E_r )}~~~~~~~~~~~~~~~~~~~~~~~~~~~~~~~~~\label{g0}  \\
 G_{\rm{1}} &=& 3 g^4 \int {d[k,q]n_k \left[ {S(E_k ,E_q ,E_r ) + S( - E_k ,E_q ,E_r )} \right]}~~\label{g1}\\
 G_{\rm{2}} &=& 3 g^4 \int {d[k,q]n_k n_q [ S(E_k ,E_q ,E_r ) + S( - E_k ,E_q ,E_r ) } + S(E_k , - E_q ,E_r ) - S(E_k ,E_q , - E_r ) ] .
 ~~\label{g2}
\eea
\end{widetext}
Here,
$S(E_k ,E_q ,E_r ) = 1/(i\omega _n  + E_k  + E_q  + E_r )+ 1/( - i\omega _n  + E_k  + E_q  + E_r )$,
$E_k E_q E_rd[k,q] = [(N+2)/24](\mu ^{4\varepsilon }/3!)[{d^{D - 1} k}/(2\pi )^{D - 1} ][{d^{D - 1} q}/(2\pi )^{D - 1} ]$,
and $r = |{\mathord{\buildrel{\lower3pt\hbox{$\scriptscriptstyle\rightharpoonup$}}
\over k}  + \mathord{\buildrel{\lower3pt\hbox{$\scriptscriptstyle\rightharpoonup$}}
\over q}  }|$.
When $N=1$, our results are compatible to those of the scalar case \cite{Parwani1992}.

\section{I\MakeLowercase{maginary} P\MakeLowercase{art} \MakeLowercase{of} $G_0$ \MakeLowercase{in} \MakeLowercase{the} L\MakeLowercase{arge} F\MakeLowercase{requency} L\MakeLowercase{imit} }

In the large frequency limit ($M(\mu = \omega) \ll \omega \ll \mu_L$),
the dominant contribution to $\Sigma''_2$
comes from ${\rm{Im}} (G_0)$ (Eq.~(\ref{g0})). Because of the on-shell energy-momentum conservations, the $\eps$ can be simply set to $0$ for calculating
 the imaginary part of the self energy with the mass term replaced by a physical one.
 Evaluating the integration for ${\rm{Im}} (G_0)$ yields
\begin{widetext}
\bea
&& {\mathop{\rm Im}\nolimits} G_0 (\omega ^2 ,\mathord{\buildrel{\lower3pt\hbox{$\scriptscriptstyle\rightharpoonup$}}
\over p} = 0) \\ \nonumber
&=& A g(\omega)^4 \pi M^2 \left[ {\int_0^{y_1 } {\frac{y}{{E_y }}\left( {\sqrt {f_4^2 (y) + 1}  - \sqrt {f_2^2 (y) + 1} } \right)dy} } \right.\left. { + \int_{y_1 }^{y_2 } {\frac{y}{{E_y }}\left( {\sqrt {f_4^2 (y) + 1}  - \sqrt {f_3^2 (y) + 1} } \right)dy} } \right] ~\label{g0imaginary}
 \eea
 \end{widetext}
where $A = (N+2)/(1152\pi^4)$, $x=k/M$, $y=q/M$, $y_1  = \sqrt {(z_0  - 1)(z_0  - 3)} /2$, $y_2  = \sqrt {(z_0^2  - 9)(z_0^2  - 1)} /(2z_0 )$
with $z_0 = \omega/M$. In addition $x=f_3(y), f_4(y)$ are two solutions of $z_0 = \sqrt{x^2+1}+\sqrt{y^2+1}+\sqrt{(x-y)^2+1}$
with $f_2(y) =  - f_3(y)$ and $f_3(y) < f_4(y)$.
In the large frequency limit we consider, $z_0 \gg 1$, $y_1 \approx z_0/2$ and $y_2 \approx y_1$, then the second integral is negligible.
In addition, using $y_1 = z_0/2 \ll z_0$,
we find ${\mathop{\rm Im}\nolimits} G_0 (\omega ^2 ,\mathord{\buildrel{\lower3pt\hbox{$\scriptscriptstyle\rightharpoonup$}}
\over p} = 0) \sim g(\omega)^4 \omega^2$.

\end{document}